\documentclass[journal]{IEEEtran}

\usepackage{cite}
\usepackage{amsmath,amssymb,amsfonts}
\usepackage{graphicx}
\usepackage{textcomp}
\usepackage{xcolor}
\usepackage{float}
\usepackage{algorithm}
\usepackage{tikz}
\usepackage{placeins}
\usepackage{algpseudocode} 
\usepackage{tabularx} 
\usepackage{booktabs}
\usepackage[utf8]{inputenc}
\usepackage{dsfont}

\algrenewcommand\algorithmiccomment[1]{\hfill\textit{// #1}}

\usetikzlibrary{arrows.meta,positioning,shapes,fit}

\def\BibTeX{{\rm B\kern-.05em{\sc i\kern-.025em b}\kern-.08em
    T\kern-.1667em\lower.7ex\hbox{E}\kern-.125emX}}
\usepackage[caption=false,font=footnotesize]{subfig}
\usepackage[hidelinks]{hyperref}
    
\begin{document}

\title{Federated Spatiotemporal Graph Learning for Passive Eavesdropping Detection on Smart Grids}

\author{Bochra~Al~Agha and Razane~Tajeddine%
\thanks{Bochra Al Agha and Razane Tajeddine are with the Department of Electrical and Computer Engineering, American University of Beirut, Beirut, Lebanon (e-mail: baa76@mail.aub.edu; razane.tajeddine@aub.edu.lb).}%
}

\maketitle
\begin{abstract}
Passive eavesdropping evades conventional intrusion detection in smart-grid wireless links because it injects no traffic and induces only subtle, asynchronous perturbations from proximity-driven propagation changes (e.g., excess shadowing and multipath/coherence reshaping). These low-amplitude, non-stationary effects are often masked by normal variability and further complicated by non-IID heterogeneity across devices, tiers (HAN/NAN/WAN), and communication technologies, while centralizing raw telemetry raises privacy and governance concerns.
This paper proposes a privacy-preserving federated spatiotemporal graph detector that jointly models neighborhood structure and short-horizon dynamics without aggregating raw data. Nodes train locally with FedProx aggregation. A graph convolutional network (GCN) encodes ego-centric topology, and a bidirectional gated recurrent unit (BiGRU) models temporal persistence over sliding windows. Inputs fuse privacy-safe engineered indicators derived from physical-layer and medium-access-control (MAC)-visible signals with topology-derived neighborhood summaries and device metadata.
Under an activity-aware, leak-safe protocol with threshold-free model selection, the proposed GCN+BiGRU attains strong detection performance at both granularities, achieving attack $F_1$ of $0.949$ at the per-timestep level and $0.937$ at the per-sequence level. Ablations show that temporal/coherence descriptors primarily drive recall, while graph context improves precision by suppressing false alarms under heterogeneous client conditions.
\end{abstract}


\section{Introduction}

Smart grids enable bidirectional communication and distributed control across heterogeneous devices and links \cite{fang2011smart}, which can expose wireless traffic to passive reconnaissance that silently reveals usage patterns, protocol metadata, and topology, thereby facilitating subsequent active attacks such as false data injection (FDIA) \cite{mo2011cyber}. Early detection of such reconnaissance mitigates FDIA risk by surfacing precursors before any manipulation occurs \cite{finster2015privacy}. Detection is challenging because observable traces are faint, transient, non-stationary, asynchronous, and inherently non-IID due to device- and context-level diversity. Moreover, practical solutions must preserve raw-data privacy at grid nodes. These challenges motivate spatiotemporal graph models that jointly encode neighborhood structure and short-horizon dynamics to improve sensitivity without inflating false positives, and that can operate under federated training constraints.

Prior work largely targets overt manipulations or assumes centralized training. Classical supervised pipelines—such as boosted trees, support vector machines (SVM), $k$-nearest neighbors ($k$-NN), artificial neural networks (ANNs), and random forests—achieve strong results on benchmark datasets or engineered features, but are most effective when attacks induce disruptive, easily separable artifacts. This limits their utility for low-signal passive attacks whose effects are weak, time-varying, and partially masked by normal variability \cite{ahmed2024hybrid,sakhnini2019smart}.

At the physical layer, many detection approaches focus on active interference or protocol/measurement manipulations, including signal injection, replay, or jamming, which produce strong channel state information (CSI) deviations but violate the assumptions of strictly passive reconnaissance \cite{prasad2019machine,hoang2021physical}. Graph-based methods show that leveraging topology can improve detection and localization; however, most operate in centralized regimes or with pooled data, raising privacy and scalability concerns for utility deployments \cite{peng2024research,jiang2024grnn}.

Federated learning (FL) trains a shared model while keeping raw measurements local and exchanging only model updates \cite{kairouz2021advances}. FL has been explored in smart grid cybersecurity, including variants for false data injection detection and node localization, showing promise for training without raw-data aggregation \cite{buyuktanir2025federated,li2022detection}. However, these studies seldom address the combination of presence-only passive reconnaissance, realistic non-IID heterogeneity, and asynchronous timing, and they rarely provide a privacy-preserving spatiotemporal graph formulation tailored to weak and transient signatures.

Consequently, a gap remains for methods that fuse graph context with short-horizon dynamics under federated learning while remaining robust to non-stationarity and enabling detection of stealthy passive reconnaissance. This work assumes a passive wireless adversary that neither injects nor modifies signals; instead, deviations arise as subtle, distributed, time-varying perturbations consistent with proximity-driven propagation changes. FL keeps data local to satisfy utility privacy and governance requirements, motivating compact encoders that learn from local time series while sharing only model updates.

\paragraph*{Problem statement and scope}
The objective is to detect presence-only passive eavesdropping on smart-grid wireless links under privacy constraints. The adversary is receive-only and does not inject, replay, or jam; detectable effects arise in proximate settings where excess shadowing and multipath/coherence reshaping induce low-amplitude deviations that are asynchronous and non-stationary. The learning setting is federated: raw telemetry remains local to nodes, clients are heterogeneous and non-IID, and the detector must remain reliable under tier (HAN/NAN/WAN) and technology diversity.

Our contributions can be summarized as follows:
\begin{itemize}
\item \textbf{Federated spatiotemporal graph detector for passive reconnaissance:}
a privacy-preserving GCN+BiGRU detector jointly encodes ego-centric neighborhood structure and short-horizon temporal persistence under FedProx aggregation without exporting raw telemetry.
\item \textbf{Leak-safe, activity-aware evaluation for low-signal attacks:}
an activity-gated reporting protocol with dual-level evaluation (per-timestep sensitivity and per-sequence operational alerting) characterizes detection under asynchronous, weak signatures.
\item \textbf{Multimodal fusion aligned with tier/technology heterogeneity:}
inputs fuse privacy-safe engineered indicators derived from physical/MAC-visible signals with topology-derived neighborhood summaries and device metadata to support analysis across HAN/NAN/WAN tiers and heterogeneous technologies.
\item \textbf{Explainability via heterogeneity analysis and ablations:}
per-client robustness is analyzed by tier and technology, and ablations isolate the contribution of temporal/coherence descriptors, spatial context, and metadata to recall and false-alarm behavior.
\end{itemize}

\section{Feature Description}
\label{sec:features}

Presence-only passive reconnaissance is modeled as a \emph{receive-only} adversary that does not transmit and performs no protocol-layer manipulation (no injection, replay, modification, jamming, routing changes, or intentional dropping). Detectability, when present, arises only because close physical proximity perturbs the propagation environment (additional absorption/diffraction causing excess shadowing, and scattering/multipath reshaping that reduces temporal coherence). This section explains \emph{how} such propagation-only perturbations translate into the exported observables and the engineered features used by the federated detector.

\subsection{Causal Justification and Threat Model}
\label{subsec:causal-justification}

\textbf{Threat model (presence-only, receive-only):}
The adversary is physically proximate (meter-scale) to an attack-eligible \emph{non-wired} link or endpoint. The model does \emph{not} assume that purely remote interception necessarily induces measurable effects; the detectable regime is explicitly the proximate one, where the adversary's body/vehicle/equipment perturbs the dominant interaction region of the link via shadowing and multipath variation \cite{3gpp38901,januszkiewicz2018bodyshadow,zhou2015multipath}. Fiber/Ethernet backbones are treated as attack-ineligible.

\textbf{Activity gating (when features can change):}
Propagation effects are only observable when legitimate transmissions occur. Therefore, perturbations and labels are applied only on active epochs (e.g., $\texttt{tx\_count}>0$). This avoids ``silent positives'' and ensures that any detectable shift is grounded in measurable link behavior rather than label artifacts.

\textbf{Channel-to-metrics dependency chain (why features change):}

To ensure physical consistency and avoid feature shortcuts, the dataset is generated through a deterministic \emph{channel-to-metrics} chain. Specifically, propagation latents first perturb the channel state information (CSI) / received signal strength indicator (RSSI)-derived observables and the effective signal-to-noise ratio (SNR) / signal-to-interference-plus-noise ratio (SINR), which then drive the packet error rate (PER) and, via retransmission/service-time inflation, latency and its exponentially weighted moving average (EWMA)-smoothed form. In proximate passive settings, excess shadowing can reduce received power, while multipath reshaping can change the exported CSI-amplitude proxy in either direction depending on the dominant path combination. Coherence degradation further increases short-horizon variability and drift statistics in CSI- and SNR-like observables. Lower SNR/SINR increases PER, while reliability and medium-access-control effects (e.g., retransmission expectation and contention/scheduling) propagate elevated PER into increased service time, queueing, and delay/jitter \cite{goldsmith2005wireless,bianchi2000performance}. Because these shifts are low-amplitude and embedded in natural variability, reliable detection benefits from (i) \emph{temporal consistency} (windowed decisions) and (ii) \emph{spatial/graph context} (neighbor consistency), rather than isolated thresholding on a single variable.

\subsection{Feature Families}
\label{subsec:feature-families}

The detector uses a leak-safe feature set derived from measurable link indicators and strictly causal descriptors. Diagnostic latent streams (e.g., internal shadowing/interference variables, if released for reproducibility) are excluded from learning inputs.

\subsubsection{Local link-indicator features (physical/MAC-visible)}
\textbf{CSI amplitude proxy:}
CSI is represented via a measurement-realistic amplitude proxy (denoted generically as $\texttt{csi\_amp}$). Presence-only perturbations (shadowing and multipath reshaping) change its distribution and short-horizon dispersion without any packet manipulation. Human proximity is widely known to induce measurable RSS/CSI shifts in device-free sensing settings \cite{patwari2010rti,palipana2016csi_presence,zhou2015multipath}.

\textbf{SNR/SINR:}
$\texttt{snr\_db}$ summarizes link-budget degradation induced by reduced received amplitude and altered effective interference superposition \cite{goldsmith2005wireless,3gpp38901}. A mild left-shift in $\texttt{snr\_db}$ under passive overlays is expected when excess shadowing is applied.

\textbf{Packet error:}
$\texttt{packet\_error}$ (PER) captures reliability effects that arise causally from SNR/SINR reduction. Under presence-only perturbations, PER changes are typically modest and intermittent, but become informative when aggregated temporally.

\textbf{Latency and smoothed latency:}
$\texttt{latency}$ and $\texttt{latency\_smooth}$ capture the downstream consequence of elevated PER via retransmission/service-time inflation and bursty delay dynamics \cite{goldsmith2005wireless,bianchi2000performance}. Smoothing (e.g., EWMA) captures short-term persistence and reduces sensitivity to single-epoch noise.

\subsubsection{Temporal/coherence descriptors (short-horizon dynamics)}
Presence-only coherence reduction and increased channel innovation manifest most reliably through \emph{rolling} and \emph{difference-based} descriptors computed strictly causally over past samples. Representative families include:
(i) \textbf{innovation/variability} statistics (e.g., rolling innovation variance of $\texttt{snr\_db}$ or $\texttt{csi\_amp}$-derived traces),
(ii) \textbf{lag structure} (e.g., lag-1 autocorrelation proxies capturing coherence loss),
and (iii) \textbf{distribution-shape spread} (e.g., quantile spread / IQR of short-horizon differences such as $\Delta \texttt{csi}$).
These descriptors target the fact that passive overlays often perturb \emph{temporal correlation} more than they shift means.

\subsubsection{Topology-derived neighbor context (spatial consistency)}
Passive windows are placed on connected local groups, so nearby nodes may exhibit \emph{co-moving} low-amplitude shifts. To exploit this, neighbor context features summarize each node relative to its 1-hop neighborhood using adjacency-weighted aggregation, producing:
(i) \textbf{neighbor averages} (e.g., $\texttt{avg\_neighbor\_snr}$, $\texttt{avg\_neighbor\_latency}$),
(ii) \textbf{deviation features} (e.g., $|\texttt{snr}-\texttt{avg\_neighbor\_snr}|$),
and (iii) \textbf{consistency/contrast descriptors} (e.g., ``rho-like'' correlation proxies, neighbor deltas, Laplacian-energy style contrast).
These features help distinguish true passive overlays (structured, locality-consistent) from isolated noise spikes.

\subsubsection{Normalization and activity masking (leak-safe usage)}
All continuous features are standardized using \emph{train-only} per-node statistics and applied to validation/test using stored parameters to prevent leakage. Training and evaluation can be restricted to active epochs or weighted by an activity mask $m(t)=\mathbb{I}[\texttt{tx\_count}(t)>0]$, reflecting that link indicators are meaningful only when traffic is present.

\subsection{Interpretation of Marginal Shifts}
\begin{figure}[t]
    \centering
    \includegraphics[width=\linewidth]{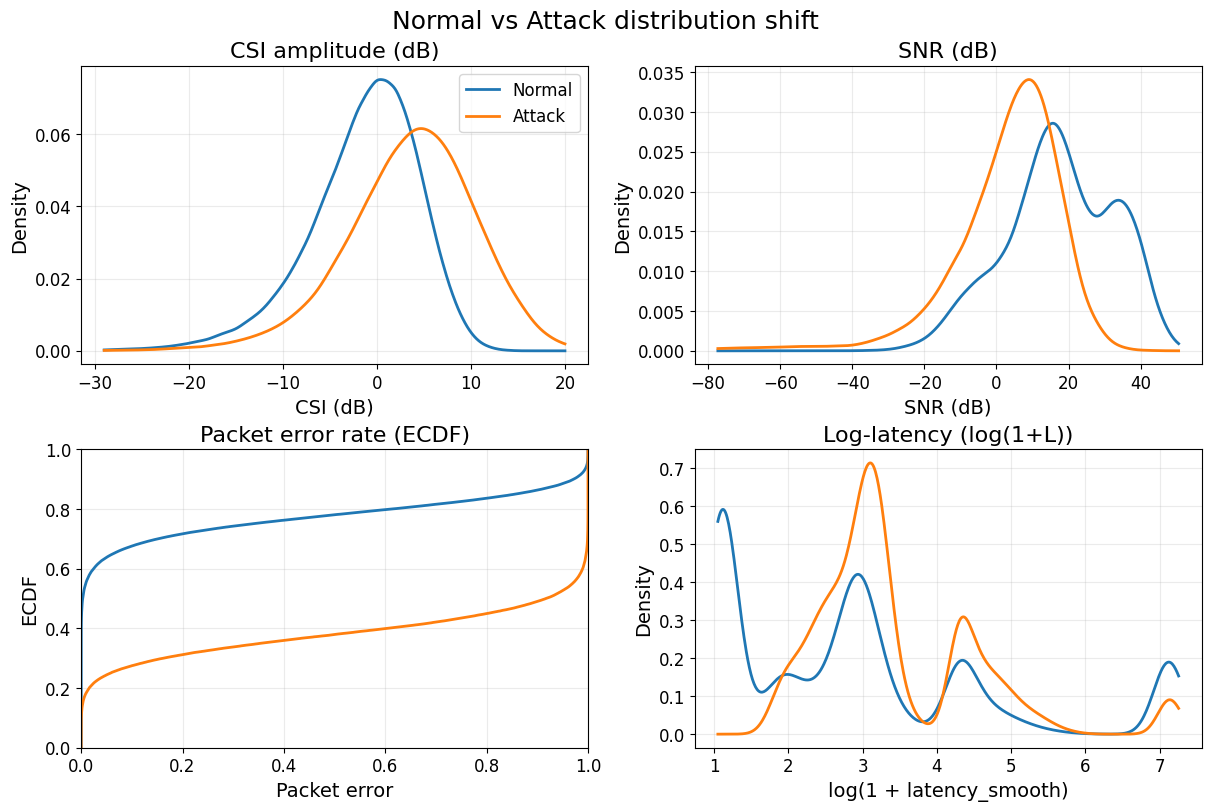}
    \caption{Marginal behavior under normal operation versus presence-only passive overlays. Kernel density estimates (KDE) are shown for CSI amplitude proxy, SNR, and $\log$-latency, and an empirical CDF (ECDF) is shown for packet error rate (PER). The strong overlap across marginals indicates a low-signal regime, motivating temporal persistence and neighborhood-consistency modeling rather than single-feature thresholding.}
    \label{fig:feature_dists}
\end{figure}

Fig.~\ref{fig:feature_dists} shows representative physical-layer and MAC-visible indicators under normal operation and presence-only passive overlays. Although strong overlap remains, consistent shifts appear across modalities: CSI amplitude exhibits tail and dispersion changes, SNR shifts slightly left, packet error increases stochastically, and smoothed latency shifts accordingly. This confirms a low-signal regime in which reliable detection benefits from temporal persistence and graph-based neighborhood consistency rather than single-feature thresholding.

% Two-column width (recommended in IEEE conference format)

\section{Federated Learning in Smart Grids}

Centralized training is often infeasible in smart grids: exporting high-rate telemetry and sensitive load profiles violates privacy and operational policies; Home, Neighborhood, and Wide Area Network links (HAN, NAN, WAN) have limited bandwidth and intermittent connectivity; devices are geographically dispersed behind utility firewalls; and client data are non-identically distributed (non-IID) \cite{zhang2024federated}. Federated learning (FL) fits these constraints by enabling on-device learning on native client distributions with partial participation, while exchanging compact model parameters/updates rather than raw payload data.

\begin{figure}[t]
    \centering
    \includegraphics[width=0.6\linewidth]{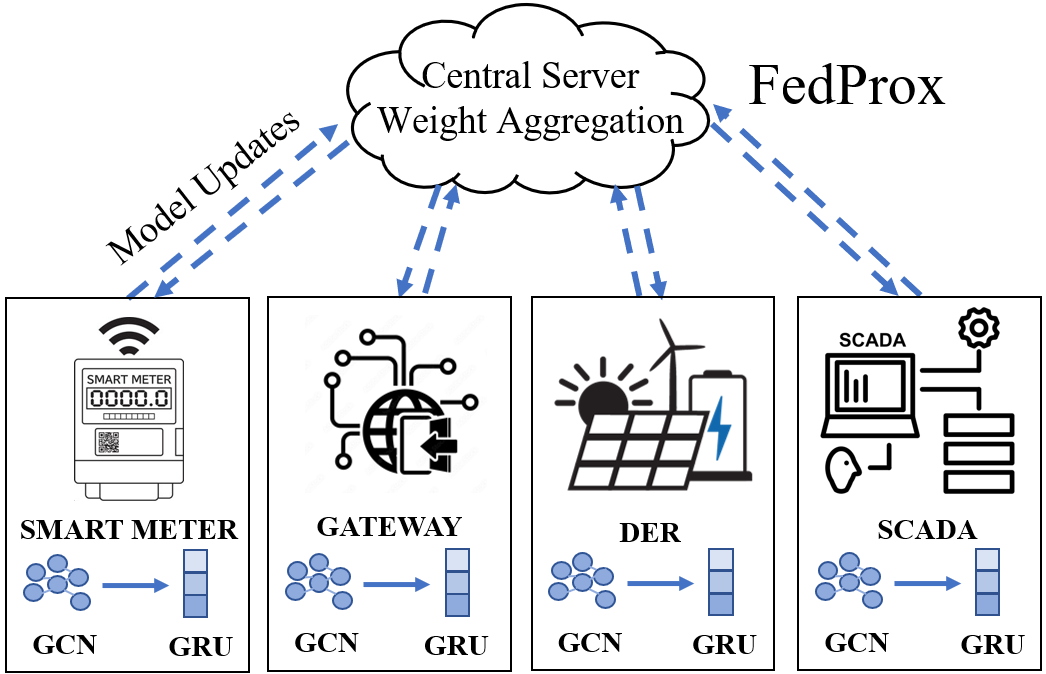}
    \caption{FedProx federated learning setup. Each device trains a local GCN+BiGRU model and sends updates; the server aggregates them by weighted averaging.}
    \label{fig:Fedprox}
\end{figure}

Fig.~\ref{fig:Fedprox} illustrates the setup: each client trains a local GCN+BiGRU detector and transmits model parameters/updates for server aggregation; the server then redistributes the updated global model. To reduce instability from client drift under heterogeneous data, Federated Proximal (FedProx) is used by adding a proximal term with coefficient $\mu$ during local optimization \cite{li2020federated}. Privacy is interpreted here as data locality; no additional privacy mechanisms (e.g., secure aggregation or differential privacy) are applied.

\section{Proposed Pipeline and Experimental Setup}
\label{sec:pipeline}

\subsection{Overall Pipeline}
\begin{figure*}[t]
    \centering
    \includegraphics[width=0.62\textwidth]{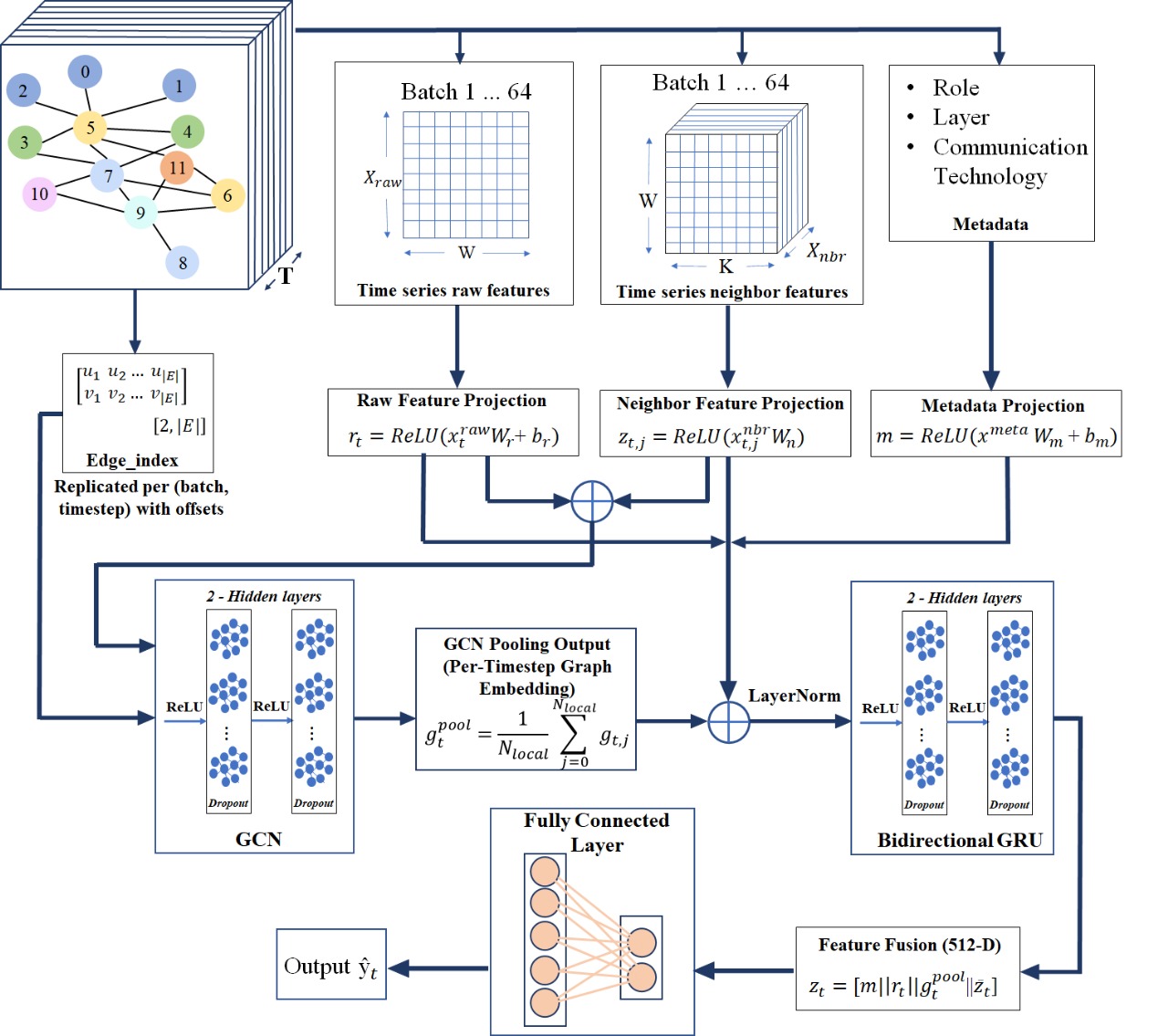}
    \caption{Federated multimodal graph-centric pipeline. Local windows are encoded with a GCN+BiGRU spatiotemporal model and aggregated using FedProx.}
    \label{fig:pipeline}
\end{figure*}

Fig.~\ref{fig:pipeline} summarizes the procedure. Multimodal features are extracted from the benchmark dataset, ego-centric star subgraphs are formed per device, and a spatiotemporal encoder applies a local graph convolutional network (GCN) for spatial context and a bidirectional gated recurrent unit (BiGRU) for temporal modeling. Training is performed in a federated manner using FedProx to stabilize optimization under heterogeneous, non-IID clients. Model selection is threshold-free and based on the validation sequence-level area under the precision-recall curve (AUPRC), after which a calibrated $(\tau,\eta)$ decision rule is applied for discrete window-level decisions.

\subsection{Graph-Structured Smart Grid and Model Inputs}
\label{sec:graph-inputs}
The smart-grid communication topology is represented as a graph with nodes as devices and edges as links, specified by an adjacency matrix $\mathbf{A}$ (or an edge index). Each wireless node acts as a federated client. For a client node $i$, model inputs over a sliding window of length $W$ are:
(i) raw per-node features $\mathbf{X}^{(\mathrm{raw})}_i\in\mathbb{R}^{W\times F_{\mathrm{raw}}}$,
(ii) topology-derived neighborhood summary features $\mathbf{X}^{(\mathrm{nbr})}_i\in\mathbb{R}^{W\times F_{\mathrm{nbr}}}$ (e.g., adjacency-weighted neighbor averages, deviations, and consistency/contrast descriptors),
and (iii) static metadata $\mathbf{m}_i\in\mathbb{R}^{F_{\mathrm{meta}}}$ (role/layer/technology indicators).

For spatial processing, each client constructs an ego-centric star subgraph
$\mathcal{G}_i^{\star}=(\{i\}\cup\mathcal{N}_i,\ \{(i,j),(j,i):j\in\mathcal{N}_i\})$
restricted to wireless neighbors $\mathcal{N}_i$. Let $K_i\triangleq|\mathcal{N}_i|$ and $N_i=1+K_i$.
In implementation, the neighborhood summary stream $\mathbf{X}^{(\mathrm{nbr})}_i$ is broadcast across the $K_i$ neighbor slots to align tensor shapes with $\mathcal{G}_i^\star$; no raw neighbor time series are exchanged across clients. Sliding windows preserve short-horizon dynamics and maintain leak-safe splits.

\subsection{Spatiotemporal Encoder (GCN + BiGRU)}
\label{sec:encoder}
For each client $i$ and time index $t$ within a window, the raw and neighborhood-summary features are projected to hidden size $H$:
\begin{align}
\mathbf{h}^{(\mathrm{raw})}_{i,t}  &= \sigma\!\left(\mathbf{W}_{r}\mathbf{x}^{(\mathrm{raw})}_{i,t}\right), \nonumber\\
\mathbf{h}^{(\mathrm{nbr})}_{i,t}  &= \sigma\!\left(\mathbf{W}_{n}\mathbf{x}^{(\mathrm{nbr})}_{i,t}\right), \label{eq:proj_emb}\\
\mathbf{h}^{(\mathrm{meta})}_{i}   &= \sigma\!\left(\mathbf{W}_{m}\mathbf{m}_{i}\right). \nonumber
\end{align}
where $\sigma(\cdot)$ denotes the rectified linear unit (ReLU) and $\mathbf{x}^{(\mathrm{raw})}_{i,t}\in\mathbb{R}^{F_{\mathrm{raw}}}$, $\mathbf{x}^{(\mathrm{nbr})}_{i,t}\in\mathbb{R}^{F_{\mathrm{nbr}}}$.

\paragraph*{Local star-graph construction.}
At each $t$, form node features in $\mathbb{R}^{2H}$ using role-split inputs:
\[
\mathbf{u}_{i,t}=[\,\mathbf{h}^{(\mathrm{raw})}_{i,t};\mathbf{0}\,],\qquad
\mathbf{u}_{j,t}=[\,\mathbf{0};\mathbf{h}^{(\mathrm{nbr})}_{i,t}\,],\ \forall j\in\mathcal{N}_i.
\]
That is, the ego node carries the projected raw stream and neighbors carry the projected neighborhood-summary stream. Two GCNConv layers (ReLU, dropout) map $2H\!\to\!2H\!\to\!H$ on $\mathcal{G}^\star_i$, yielding embeddings for all $N_i$ nodes. The spatial representation used for fusion is the \emph{ego} embedding after the second GCN layer:
\[
\mathbf{g}_{i,t}\in\mathbb{R}^{H}.
\]

\paragraph*{Fusion and temporal modeling.}
In parallel, define the neighborhood summary embedding $\bar{\mathbf{h}}^{(\mathrm{nbr})}_{i,t}\!=\!\mathbf{h}^{(\mathrm{nbr})}_{i,t}$ (equivalently the mean over broadcast neighbor slots), keep $\mathbf{h}^{(\mathrm{raw})}_{i,t}$, and include $\mathbf{h}^{(\mathrm{meta})}_{i}$.
The fused token is
\[
\mathbf{z}_{i,t}=\mathrm{LN}\!\Big([\mathbf{g}_{i,t}\|\bar{\mathbf{h}}^{(\mathrm{nbr})}_{i,t}\|\mathbf{h}^{(\mathrm{meta})}_i\|\mathbf{h}^{(\mathrm{raw})}_{i,t}]\Big)\in\mathbb{R}^{4H},
\]
where $\|\,$ denotes concatenation and LN is feature-wise layer normalization. The sequence $\{\mathbf{z}_{i,t}\}_{t=1}^{W}$ is fed to a 2-layer bidirectional GRU with hidden size $H_{\mathrm{gru}}$, followed by a linear head producing per-timestep logits $\boldsymbol{\ell}_{i,t}=[\ell^{(0)}_{i,t},\ell^{(1)}_{i,t}]$. The per-timestep attack probability is
\[
p_{i,t}=\frac{e^{\ell^{(1)}_{i,t}}}{e^{\ell^{(0)}_{i,t}}+e^{\ell^{(1)}_{i,t}}}.
\]

\subsection{Loss Function}
\label{sec:loss}
Let $B$ denote the mini-batch size (number of windows), and let $W$ denote the window length. For window $i\in\{1,\dots,B\}$ and timestep $t\in\{1,\dots,W\}$, let $y_{i,t}\in\{0,1\}$ be the timestep label and let $a_{i,t}\in\{0,1\}$ be an activity mask indicating timesteps with legitimate traffic (active epochs). Let $(w_0,w_1)$ denote class weights computed per client from its (masked) training data. The masked, class-weighted per-timestep cross-entropy is
\[
\mathcal{L}_{\mathrm{t}}
=\frac{1}{\sum_{i=1}^{B}\sum_{t=1}^{W} a_{i,t}}
\sum_{i=1}^{B}\sum_{t=1}^{W}
a_{i,t}\,\mathrm{CE}\!\big(\boldsymbol{\ell}_{i,t},y_{i,t};w_0,w_1\big).
\]

To align training with window-level detection, define the sequence label
\[
y^{\mathrm{seq}}_{i}=\mathds{1}\!\left[\sum_{t=1}^{W} a_{i,t}\,y_{i,t}>0\right],
\]
and define active timestep probabilities $p^{\mathrm{act}}_{i,t}=a_{i,t}\,p_{i,t}$. Let $\mathcal{T}_k(i)$ be the indices of the top-$k$ values of $\{p^{\mathrm{act}}_{i,t}\}_{t=1}^{W}$ (with $k\le W$).In all experiments, $k=3$. The sequence probability is computed using a top-$k$ noisy-OR:
\[
\tilde{p}_i = 1-\prod_{t\in\mathcal{T}_k(i)}(1-p^{\mathrm{act}}_{i,t}),
\]
and the sequence loss is binary cross-entropy,
\[
\mathcal{L}_{\mathrm{seq}}
=-\frac{1}{B}\sum_{i=1}^{B}\Big[y^{\mathrm{seq}}_{i}\log \tilde{p}_i+(1-y^{\mathrm{seq}}_{i})\log(1-\tilde{p}_i)\Big].
\]
The supervised local objective is
\[
\mathcal{L}_{\mathrm{sup}}=\alpha\,\mathcal{L}_{\mathrm{t}}+\lambda_{\mathrm{seq}}\,\mathcal{L}_{\mathrm{seq}},
\]
where $\alpha=0.7$ and $\lambda_{\mathrm{seq}}=0.20$. Weight decay and gradient clipping are used for stability. Threshold $\tau$ and persistence parameter $\eta$ (defined in Sec.~\ref{sec:calibration}) are \emph{not} used during training; they are calibrated on validation after checkpoint selection and applied only at evaluation.

\subsection{Decision Rule and Calibration ($\tau$, $\eta$)}
\label{sec:calibration}
Training and checkpoint selection are threshold-free: the global checkpoint is chosen using validation sequence AUPRC without committing to a probability threshold. After selecting the best checkpoint, a discrete decision rule is calibrated on the validation set by grid search over $(\tau,\eta)$.

Given per-timestep probabilities $p_{i,t}$ and activity mask $a_{i,t}$, a window-level decision is produced by thresholding and enforcing an $\eta$-of-$W$ persistence rule over active timesteps:
\[
\hat{y}^{\mathrm{seq}}_i(\tau,\eta)=\mathds{1}\!\left[\sum_{t=1}^{W} a_{i,t}\,\mathds{1}(p_{i,t}>\tau)\ \ge\ \eta\right],
\]
where $\tau\in(0,1)$ is the probability threshold and $\eta\in\{1,\dots,W\}$ is the minimum number of active timesteps exceeding $\tau$ required to flag the window. Calibration pools validation windows across all clients, searches $\tau\in\{0.05,0.06,\dots,0.95\}$ and $\eta\in\{1,\dots,W\}$, and selects $(\tau^\star,\eta^\star)$ to maximize the attack-class $F_1$. Ties are broken by higher attack recall, then higher attack precision, then lower false-positive rate, and finally smaller $\eta$. The calibrated $(\tau^\star,\eta^\star)$ is then fixed and applied unchanged to test data to avoid leakage.

\subsection{Federated Learning Framework}
\label{sec:fl-framework}
Each wireless node acts as an FL client. Let $M$ denote the number of participating clients. A central server runs $R$ federated rounds by broadcasting global parameters, collecting local parameter updates, and aggregating them. Raw data remain local.

\paragraph*{Non-IID and heterogeneous clients}
Clients differ in roles, layers, technologies, and neighborhood structure, inducing non-identically distributed data:
\[
P_{\mathcal{D}_1} \neq P_{\mathcal{D}_2} \neq \cdots \neq P_{\mathcal{D}_M},
\]
where $\mathcal{D}_i$ is client $i$’s dataset.

\paragraph*{FedProx local objective}
Let $\theta^{(g)}_r$ denote the broadcast global parameters at round $r$ and $\theta_i$ the locally optimized client parameters returned by client $i$. FedProx adds a proximal regularizer during local training:
\[
\min_{\theta}\ \mathcal{L}_{\mathrm{sup}}(\theta)\;+\;\frac{\mu}{2}\|\theta-\theta^{(g)}_r\|_2^2,
\]
where $\mu>0$ reduces client drift under heterogeneity \cite{li2020federated}. After local training, selected clients $\mathcal{S}_r\subseteq\{1,\dots,M\}$ return parameters $\theta_i$ with weights $n_i$ (e.g., number of local training windows); the server aggregates by weighted averaging:
\begin{equation}
\label{eq:fedavg}
\theta^{(g)}_{r+1} \leftarrow \frac{\sum_{i\in\mathcal{S}_r} n_i\,\theta_i}{\sum_{i\in\mathcal{S}_r} n_i}.
\end{equation}
Setting $\mu=0$ recovers federated averaging (FedAvg). Experiments use $R=10$ communication rounds; each selected client performs one local epoch per round unless otherwise stated (Sec.~\ref{sec:expt-setup}).

\section{Experimental Setup}
\label{sec:expt-setup}

\subsection{Benchmark Overview}
The dataset used in this study is described in \cite{agha2026benchmarkingdatasetpresenceonlypassive}. The corresponding implementation and configuration files are publicly available on GitHub.\footnote{\url{https://github.com/bochraagha/smartgrid-passive-attack-dataset-generator}}

\subsection{Data Splitting and Leakage Prevention}
Train/validation/test splits follow a leak-safe protocol based on independent realizations (split-specific seeds) with burn-in removal. All feature engineering is strictly causal within each split, and per-node normalization is fitted on the training split only and reused for validation/test. Windowing is performed within each split, and calibration of post-hoc decision thresholds is performed on validation only.

\subsection{Implementation Details}
\label{sec:impl}
The implementation of the proposed federated GCN+BiGRU detector is publicly available on GitHub.\footnote{\url{https://github.com/bochraagha/federated-gcn-gru-passive-attack-detection}} The proposed method is implemented in PyTorch with Flower.Each wireless node acts as a federated client and trains on its ego-star subgraph using mini-batches of 64 windows with window length $W{=}9$. The encoder consists of two GCNConv layers with ReLU activations (hidden width $H{=}128$), followed by a 2-layer bidirectional GRU with 192 units per direction. Dropout is set to 0.2 after the GCN and within the GRU. Training uses Adam with learning rate $1{\times}10^{-3}$ and weight decay $5{\times}10^{-5}$, and gradients are clipped to $\lVert\nabla\theta\rVert_2\!\le\!1.0$.

Federated training runs for $R{=}10$ communication rounds with client fraction 1.0 (all available wireless clients) and FedProx regularization ($\mu{=}0.01$). The supervised objective combines a masked timestep loss (weight $\alpha{=}0.7$) with an auxiliary sequence-level ``any-attack'' loss (weight $\lambda_{\mathrm{seq}}{=}0.20$). Class weights are computed per client using active timesteps only when activity masking is enabled.

Model selection is \emph{threshold-free}: during training, the best global checkpoint is selected by validation sequence-level AUPRC computed from the maximum active timestep probability within each window. This score is used only for checkpoint selection; the auxiliary sequence loss during training still follows the top-$k$ noisy-OR formulation defined in Sec.~\ref{sec:loss}. Reported results include both per-timestep and per-sequence metrics (precision, recall, $F_1$, and false-positive rate), computed using the calibrated decision rule in Sec.~\ref{sec:calibration}.

\paragraph*{Feature inputs}
Per-window inputs include raw z-scored features (e.g., \texttt{z\_snr\_db}, \texttt{z\_packet\_error}, and \texttt{z\_latency\_smooth}), together with temporal innovation, autocorrelation-function (ACF), and spread descriptors, topology-derived neighborhood summary features (adjacency-weighted neighbor averages, deviation/contrast measures, and ``rho-like'' consistency terms), and one-hot encoded device metadata (role, layer, communication technology, and link type). All features are standardized using training-only statistics and applied unchanged to validation/test.

\paragraph*{Inference rule and calibration}
Inference uses the calibrated $(\tau^\star,\eta^\star)$ obtained by global validation grid search as described in Sec.~\ref{sec:calibration}, and is applied unchanged at test time to avoid leakage.

\section{Results and Discussion}
\subsection{Evaluation Protocol and Metrics}
\label{subsec:eval_protocol_metrics}

All reported results follow the activity-aware, leak-safe evaluation protocol specified in Sec.~\ref{sec:expt-setup} and the calibrated decision rule in Sec.~\ref{sec:calibration}. Performance is summarized at two complementary granularities to reflect (i) instantaneous detectability and (ii) operational alerting over short horizons.

\textbf{Per-timestep (epoch-level) evaluation:}
each \emph{active} epoch is treated as a binary decision derived from the per-timestep attack probability. Metrics are computed only over active epochs, consistent with the activity filtering defined earlier, and reported using attack-class precision, recall, $F_1$, and false-positive rate.

\textbf{Per-sequence (window-level) evaluation:}
each window is mapped to a single binary label under the window labeling rule already defined in Sec.~\ref{sec:loss}. A window-level prediction is produced by applying the calibrated persistence rule $(\tau^\star,\eta^\star)$ from Sec.~\ref{sec:calibration}, which enforces temporal consistency by requiring multiple above-threshold epochs within the same window. Window-level metrics are reported using the same attack-class measures.

\textbf{Model selection and calibration:}
checkpoint selection is threshold-free (validation sequence AUPRC), after which $(\tau^\star,\eta^\star)$ is tuned on validation and then fixed for test evaluation, as described in Sec.~\ref{sec:calibration}. This separation ensures that test-time reporting uses a single frozen rule rather than threshold tuning on the test set.

To characterize heterogeneity under non-IID clients, results are reported both as global aggregates and as per-client breakdowns in subsequent subsections.

\subsection{Overall Detection Performance}
\label{subsec:overall_detection}

This subsection reports the test performance of GCN+BiGRU using the frozen calibration rule described earlier. Results are presented at both granularities to separate instantaneous detectability (per-timestep) from operational alerting over windows (per-sequence).

\textbf{Per-timestep (GLOBAL):} the model achieves accuracy $0.9780$ with attack-class precision $0.9838$, recall $0.9166$, and $F_1$ $0.9490$. Normal-class recall remains high ($0.9957$), indicating that normal active epochs are preserved with minimal confusion.

\textbf{Per-sequence (GLOBAL):} the model achieves accuracy $0.9766$ with attack-class precision $0.9846$, recall $0.8935$, and $F_1$ $0.9369$. The high attack precision indicates that window-level alerts are rarely triggered spuriously under the persistence constraint, while the recall reduction relative to per-timestep performance is consistent with stricter window-level evidence requirements: windows dominated by borderline or short-lived deviations may not accumulate enough above-threshold active epochs to satisfy the persistence rule. This separation between per-timestep and per-sequence behavior provides interpretability: per-timestep metrics quantify sensitivity to weak instantaneous signatures, whereas per-sequence metrics quantify robustness of actionable alerts under temporally consistent evidence.

\begin{figure*}[t]
    \centering
    \subfloat[Per-timestep attack precision/recall by client.]{
        \includegraphics[width=0.48\textwidth]{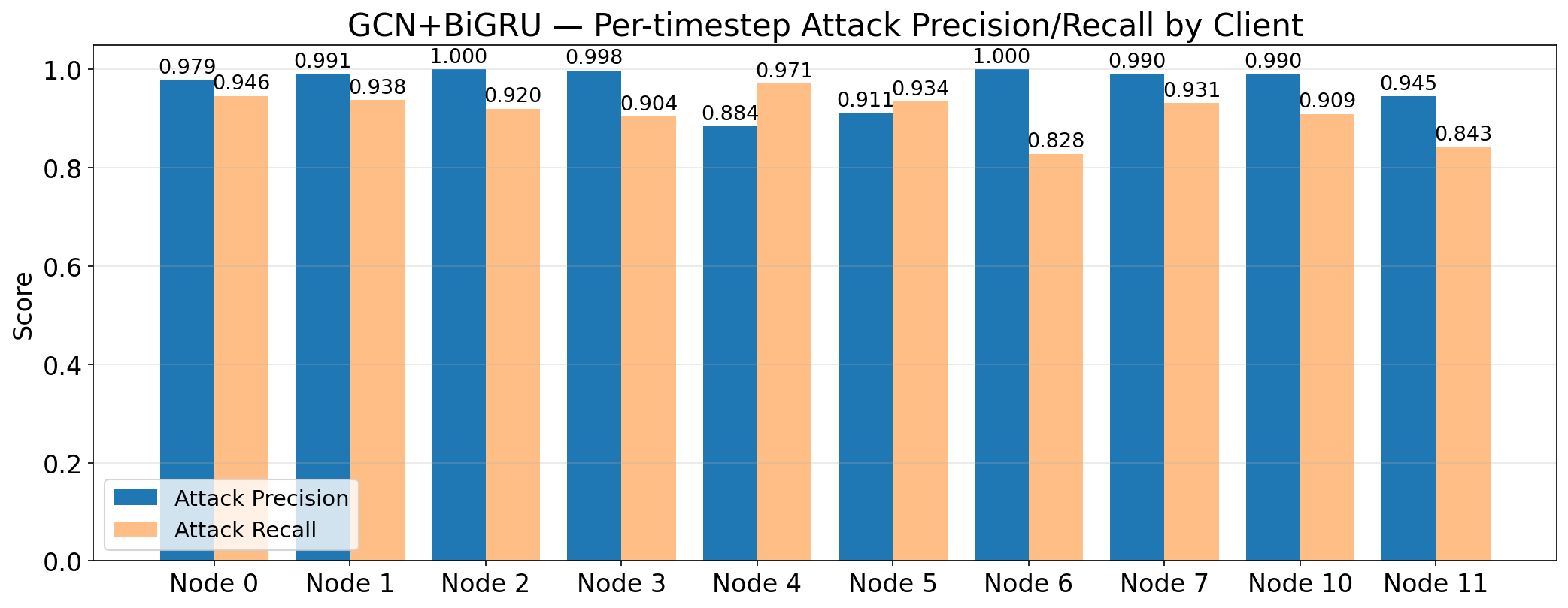}
        \label{fig:client_pr_timestep}
    }\hfill
    \subfloat[Per-sequence attack precision/recall by client.]{
        \includegraphics[width=0.48\textwidth]{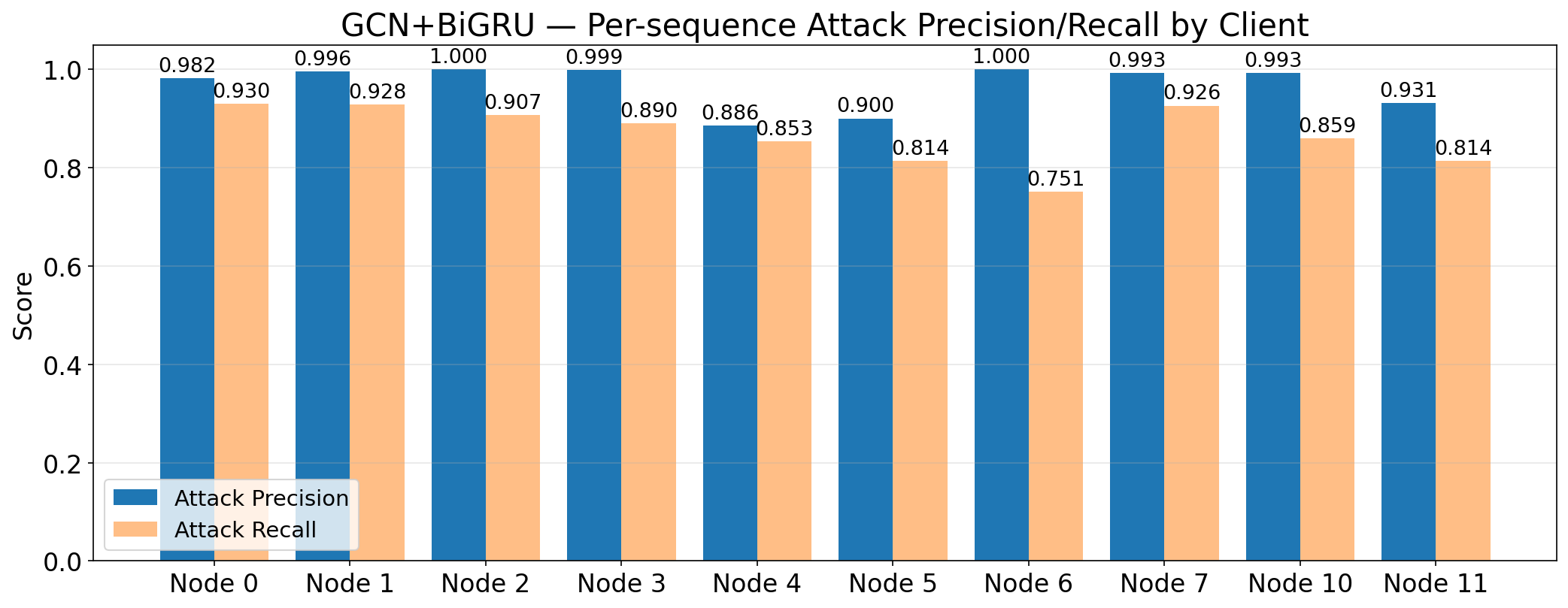}
        \label{fig:client_pr_sequence}
    }
    \caption{Per-client attack precision/recall for GCN+BiGRU (ours) under non-IID client data. Per-timestep metrics quantify epoch-level sensitivity, while per-sequence metrics reflect window-level alerts under the calibrated persistence rule. Nodes 8 and 9 are excluded because the attack class is absent in the corresponding test partitions, making attack recall undefined.}
    \label{fig:client_pr}
\end{figure*}

% ============================================================
\subsection{Per-Client Robustness and Heterogeneity}
\label{subsec:per_client}

Client-level performance reflects heterogeneous roles, layers, and communication technologies, resulting in non-identically distributed operating regimes. Fig.~\ref{fig:client_pr_timestep} and Fig.~\ref{fig:client_pr_sequence} report attack-class precision and recall per client for GCN+BiGRU (ours), excluding Nodes 8 and 9 since both correspond to fiber links and are attack-ineligible in the benchmark metadata.

Across clients, attack precision remains consistently high. At the per-timestep level (Fig.~\ref{fig:client_pr_timestep}), precision ranges from $0.884$ to $1.000$, indicating that attack decisions are rarely triggered spuriously on active epochs. At the per-sequence level (Fig.~\ref{fig:client_pr_sequence}), precision remains similarly strong ($0.886$ to $1.000$), showing that window-level alerts remain conservative under the fixed persistence rule.

In contrast, attack recall exhibits meaningful heterogeneity. Per-timestep recall spans $0.828$ to $0.971$ (Fig.~\ref{fig:client_pr_timestep}), while per-sequence recall spans $0.751$ to $0.930$ (Fig.~\ref{fig:client_pr_sequence}). The systematic recall reduction from per-timestep to per-sequence evaluation is consistent with the persistence constraint: windows containing short-lived or borderline-evidence deviations may not accumulate enough above-threshold active epochs to satisfy the calibrated $\eta$-of-$W$ rule. The largest recall drops occur at Node~4 ($0.971\rightarrow0.853$) and Node~5 ($0.934\rightarrow0.814$), suggesting that many attacked windows on these clients exhibit intermittent evidence rather than sustained deviations throughout the window.

Precision remains high across tiers and technologies, while recall is more sensitive to how persistently attack evidence appears within a window. HAN, ZigBee, and Wi-Fi clients show the strongest per-sequence recall, whereas NAN, LoRa, and PLC settings exhibit larger drops, indicating more intermittent evidence under the calibrated persistence rule. Notably, the LoRa DER clients (Nodes 4 and 5) retain high per-timestep recall but show lower sequence-level recall, suggesting that many attacked windows contain only a few high-confidence active epochs.

% ============================================================
\begin{figure*}[t]
    \centering
    \subfloat[GLOBAL per-timestep attack precision/recall.]{
        \includegraphics[width=0.48\textwidth]{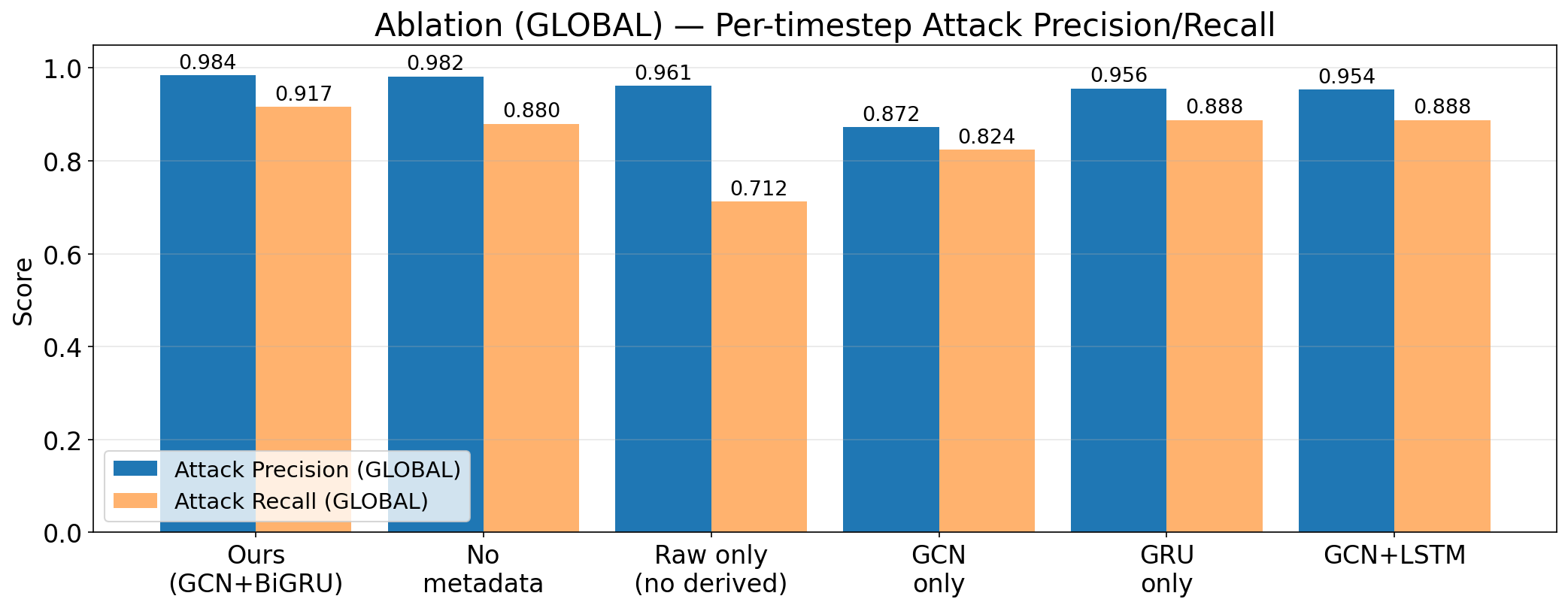}
        \label{fig:ablation_pr_timestep}
    }\hfill
    \subfloat[GLOBAL per-sequence attack precision/recall.]{
        \includegraphics[width=0.48\textwidth]{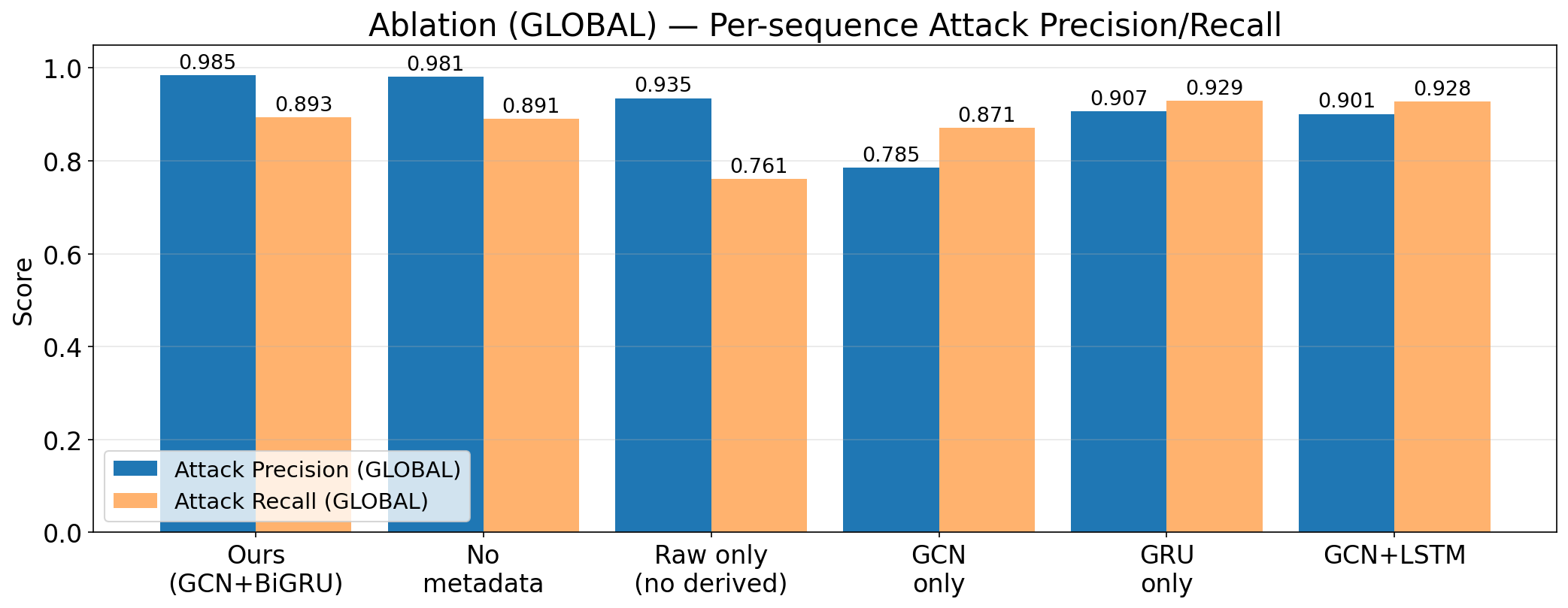}
        \label{fig:ablation_pr_sequence}
    }
    \caption{Global ablation study under the same evaluation protocol (activity-aware reporting and validation-calibrated persistence). Per-timestep metrics quantify epoch-level sensitivity, while per-sequence metrics reflect window-level alerting under the calibrated $(\tau^\star,\eta^\star)$ rule.}
    \label{fig:ablation_global}
\end{figure*}

% ============================================================
\subsection{Ablation Study}
\label{subsec:ablation}

This subsection isolates the contributions of (i) engineered temporal/coherence descriptors, (ii) graph-based spatial context, (iii) static metadata, and (iv) the temporal backbone choice. Fig.~\ref{fig:ablation_pr_timestep}--Fig.~\ref{fig:ablation_pr_sequence} report global attack-class precision and recall at both granularities.

\textbf{Full model (GCN+BiGRU).}
The complete configuration achieves per-timestep precision/recall $0.984/0.917$ (attack $F_1\approx0.949$) and per-sequence precision/recall $0.985/0.893$ (attack $F_1\approx0.937$). The resulting operating point maintains near-maximal precision while preserving strong recall under the window-level persistence rule.

\textbf{Removing metadata.}
Dropping metadata primarily impacts fine-grained detection: per-timestep recall decreases from $0.917$ to $0.880$ (attack $F_1\approx0.928$), indicating that role/layer/technology descriptors help disambiguate borderline active epochs whose local features overlap with normal variability. In contrast, per-sequence performance remains close to the full model ($0.981/0.891$, attack $F_1\approx0.934$), consistent with window-level persistence and multi-modal temporal features already providing redundant evidence at the alerting timescale.

\textbf{Raw features only (no derived descriptors).}
Using raw observables without derived temporal/coherence statistics produces the largest degradation, particularly in recall: per-timestep precision/recall drops to $0.961/0.712$ (attack $F_1\approx0.818$) and per-sequence to $0.935/0.761$ (attack $F_1\approx0.839$). This behavior is consistent with the strong marginal overlap observed under presence-only passive overlays: engineered short-horizon descriptors (innovation/dispersion/lag structure) provide the separability required to surface weak, transient deviations beyond mean-shift cues.

\textbf{Spatial-only (GCN only).}
A spatial model without temporal memory yields markedly lower precision at the window level ($0.785$) despite moderate recall ($0.871$), indicating elevated false alarms when temporal consistency is not explicitly modeled. At the timestep level, the same trend appears as reduced precision ($0.872$) relative to temporally-aware variants. These results indicate that topology alone is insufficient to stabilize detection under non-stationarity and asynchronous effects; temporal modeling is needed to suppress isolated graph-consistent fluctuations that are not persistently attack-driven.

\textbf{Temporal-only (GRU only).}
A purely temporal model attains strong recall, especially per-sequence ($0.929$), but with lower precision ($0.907$) than the full model (attack $F_1\approx0.918$). This suggests that temporal persistence captures much of the attack evidence, while spatial context primarily reduces false positives by enforcing neighborhood-consistent structure and separating localized passive overlays from client-specific noise bursts.

\textbf{Temporal backbone choice (GCN+LSTM vs. GCN+BiGRU).}
Replacing BiGRU with LSTM reduces precision while maintaining similar recall: per-sequence precision/recall is $0.901/0.928$ (attack $F_1\approx0.914$). The BiGRU variant yields a more conservative and better-calibrated alerting behavior (higher precision at comparable recall), improving the overall precision--recall trade-off under the fixed persistence rule.

Overall, the ablation confirms that derived temporal/coherence descriptors provide the dominant recall gains for low-signal passive effects, spatial context primarily increases precision by suppressing false alarms, and metadata contributes most at the fine-grained (per-timestep) level where client heterogeneity is most pronounced.

\section{Conclusion}
\label{sec:conclusion}

This paper presents a privacy-preserving detector for presence-only passive eavesdropping in smart-grid wireless links using federated spatiotemporal graph learning. The method combines local graph context (GCN) with short-horizon temporal modeling (BiGRU) and an activity-aware evaluation protocol with a validation-calibrated persistence rule for window-level alerting. Under the proposed protocol, GCN+BiGRU achieves strong global performance in a low-signal regime, with per-timestep attack precision/recall of $0.9838/0.9166$ ($F_1{=}0.9490$) and per-sequence attack precision/recall of $0.9846/0.8935$ ($F_1{=}0.9369$), indicating conservative alerting with high reliability under temporally consistent evidence.

Client-level analysis further highlights robustness under non-IID heterogeneity: attack precision remains high across clients, while recall varies in a manner aligned with tier and technology. The persistence-induced recall reduction is smallest in tiers/technologies where evidence is sustained across active epochs, and largest where attack signatures are intermittent within windows. Ablation results confirm that engineered temporal/coherence descriptors drive the dominant recall gains in the presence-only setting, spatial context primarily increases precision by suppressing false alarms, and metadata contributes most at the fine-grained (per-timestep) level where client heterogeneity is most pronounced. Future work includes extending evaluation to additional topologies and real deployments, incorporating stronger privacy mechanisms (e.g., secure aggregation or differential privacy), and expanding the threat model toward joint passive/active scenarios and localization-oriented outputs.

\bibliographystyle{IEEEtran}
\bibliography{references}
\end{document}